\documentclass[12pt]{iopart}
\usepackage{graphicx}
\usepackage{amsmath}
\usepackage{amssymb}
\usepackage{color}
\usepackage{caption}

\newcommand{\stella}{\texttt{stella}}
\newcommand{\blue}[1]{\textcolor{black}{#1}}
\newcommand{\revone}[1]{\textcolor{black}{#1}}
\newcommand{\revtwo}[1]{\textcolor{black}{#1}}
\newcommand{\refone}[1]{\textcolor{black}{#1}}
\newcommand{\reftwo}[1]{\textcolor{black}{#1}}

\begin{document}

\title[Turbulent transport of impurities in 3D devices]{Turbulent transport of impurities in 3D devices}

\author{J.~M.~Garc\'ia-Rega\~na$^1$, M. Barnes$^2$, I. Calvo$^1$,\\ A. Gonz\'alez-Jerez$^1$, H. Thienpondt$^1$, E. S\'anchez$^1$, F. I. Parra$^2$ and D. A.  St.-Onge$^2$}

\address{$^1$Laboratorio Nacional de Fusi\'on-CIEMAT, Avda. Complutense 40, 28040, Madrid, Spain\\
	$^2$Rudolf Peierls Centre for Theoretical Physics, University of Oxford, Oxford OX1 3NP, United Kingdom}
\ead{jose.regana@ciemat.es}
\vspace{10pt}
\begin{indented}
\item[]Received 31 May 2021, revised 21 July 2021
\item[]Accepted for publication 13 August 2021
\item[]Published 29 September 2021
\end{indented}

\begin{abstract}
The evidence of a large diffusive turbulent contribution to the radial impurity transport 
in Wendelstein 7-X (W7-X) plasmas has been experimentally inferred during the first campaigns 
and numerically confirmed by means of gyrokinetic simulations with the code \stella. 
In general, the absence of strong impurity accumulation during the initial W7-X campaigns  
is attributed to this diffusive term. 
Given the large variety of possible stellarator configurations, in the present work 
the diffusive contribution is also calculated in other stellarator plasmas. In particular, 
a numerical cross-device comparison is presented, where the diffusion ($D$) and convection ($V$) coefficients
of carbon and iron impurities produced by \revtwo{ion-temperature-gradient} (ITG) turbulence are obtained. 
The simulations have been performed for
the helias W7-X, the heliotron LHD, the heliac TJ-II and the quasi-axisymmetric stellarator NCSX \refone{at the radial position $r/a=0.75$}.
The results show that, although the size of $D$ and $V$ can differ across the four devices, inward convection is found for all of them. For W7-X, TJ-II and NCSX the two coefficients are comparable and the turbulent peaking factor is surprisingly similar. 
In LHD, appreciably weaker diffusive and convective impurity transport and  significantly larger turbulent peaking factor, in comparison with the other three stellarators, are predicted. All this suggests that ITG turbulence, although not strongly, would lead to negative impurity density gradients in stellarators. 
Then, considering mixed ITG/Trapped Electron Mode (TEM) turbulence for the specific case of W7-X, it has been quantitatively assessed to what degree pellet fueled reduced turbulence scenarios feature reduced turbulent transport of impurities as well. The results for trace iron impurities show that, although their turbulent transport is not entirely suppressed, a significant reduction of the convection and a stronger decrease of the diffusion term are found. Although the diffusion is still above neoclassical levels, the neoclassical convection would gain under such conditions a greater specific weight on the dynamics of impurities in comparison with \refone{standard ECRH scenarios without pellet fueling}. 
\end{abstract}

%

%
%
%
%

\section{Introduction}

Impurity transport has been a very active area of research in plasma physics due to the limitations that high core impurity concentration imposes to achieving high performance conditions in future fusion reactors. In stellarators, where neoclassical transport is larger than in tokamaks, the theoretical and numerical \revone{study} of impurity transport has been practically limited to the framework of neoclassical theory. In general, the sign of the neoclassical radial electric field, negative (ion root) or positive (electron root), correlates well with long and short, respectively, impurity confinement time. The neoclassical formalism has recently undergone an intense revision and included ingredients traditionally neglected, see e.g.~refs.~\cite{Regana_nf_57_056004_2017, Calvo_ppcf_59_055014_2017, Calvo_jpp_2018, Calvo_NF_58_124005_2018, Buller_jpp_84_2018, Mollen_ppcf_60_084001_2018,  Velasco_ppcf_60_07400_2018, Velasco_jcp_2020}.
These extended formulations and numerical approaches have shown important corrections to the impurity transport predicted by the standard neoclassical theory and provided  explanations of experimental scenarios that contradicted the standard framework, like the impurity hole LHD conditions \cite{Ida_pop_16_056111_2009,Fujita_nf_submitted_2021}. \refone{Also, the importance of the classical transport channel has been revisited and confirmed for W7-X \cite{Buller2019}.}
However, recent experiments in W7-X have routinely found that the transport of impurities have distinctive features that point out to an important role of turbulence, in particular through its diffusive contribution \cite{Geiger_NF_59_046009_2019}. The absence of impurity accumulation in most scenarios during the first experimental W7-X campaigns has been frequently attributed to this large diffusive turbulent transport source \cite{Klinger_nf_59_112004_2019}.
Among the investigations on the transport of impurities in W7-X that support this statement, the following are worth detailing. On the experimental side, the lack of appreciable charge state dependence of the impurity confinement time, $\tau_I$, has been confirmed in W7-X plasmas \cite{Langenberg2020}. In addition, $\tau_I$ has been shown to become longer, and the diffusion coefficient ($D$) weaker, with increasing ion to electron temperature ratio, $T_i/T_e$, which stabilizes the \revone{linear} ion temperature gradient (ITG) instability \cite{Wegner_NF_2020}. In pellet-fueled enhanced confinement \revone{scenarios}, where ion-root conditions are achieved \blue{throughout the plasma core \cite{Pablant_nf_2020}} and turbulent transport and fluctuations are reduced, the impurity confinement time also increases \cite{Stechow_submitted_2021} and the impurity density profile develops strong gradients at radial core locations \cite{Langenberg_IAEA_2021}. Finally, on the numerical side, the first nonlinear gyrokinetic simulations including kinetic impurities, apart from kinetic main ions and electrons, have been performed for W7-X \cite{Regana_JPP_2021} with the code \texttt{stella} \cite{Barnes_jcp_391_2019}. In particular, the calculations presented in \cite{Regana_JPP_2021}
considered different impurities at trace concentration and, for density-gradient-driven trapped electron mode (TEM) and ITG turbulence, obtained the diffusion coefficient ($D_{Z1}=D$), the thermo-diffusion coefficient ($D_{Z2}$) and the convection coefficient ($C_{Z}$) in the absence of an impurity pressure gradient. The weak dependence of the three coefficients on the impurity charge was confirmed. Regarding $D_{Z2}$ and $C_Z$, both were shown to contribute to drive impurities radially inwards for the ITG case (not for TEM turbulence, where $C_Z$ drives a weak anti-pinch contribution). With respect to $D_{Z1}$, its large size was found to be compatible with the experimentally inferred values and far above the neoclassical \revone{predictions}. It was also shown that $D_{Z1}$ is larger for ITG than for TEM turbulence, although the TEMs were appreciably 
more unstable linearly than the ITG instability. However, 
for comparison purposes, equal values for the normalised gradient scale length driving each case 
were considered, $a/L_{T_i}=4$ for the ITG case and $a/L_{n_e}=4.0$ for the TEM case \refone{(for a given quantity $Q$, $a/L_{Q}=-a(\mathrm{d}\ln Q/\mathrm{d}r)$, with $a$ the minor radius of the device and $r$ the flux surface label)}. This, together with the fact that W7-X plasmas rarely reach such 
large density gradients draws the conclusion that, in general, the turbulent part of the impurity transport will be 
dominated by ITG turbulence rather than by TEM turbulence.

Thus, in the light of the \revone{experimental and numerical results} just summarized, 
which stresses the role of ITG turbulence on the behavior of impurity transport in W7-X, it is natural to wonder: (i) whether ITG-driven impurity transport has a particular character in W7-X or has features in common with other 
stellarator configurations;  (ii) and given the strong interest in the high performance scenarios of W7-X, to what degree the density peaking of the main species, which reduces the background turbulence levels, leads also to a reduction of the turbulent impurity transport.
The remainder of the paper is organized as follows. In section \ref{sec:turbulence} the \revtwo{linear} stability and turbulence of a pure ITG setting (i.e.~that is driven solely by the ion temperature gradient with all other gradients set to zero) is studied for different stellarators. In particular, the helias W7-X, the heliotron LHD, the heliac TJ-II and the quasi-axisymmetric stellarator NCSX
are compared. 
In section \ref{sec:transport}, question (i) above, about the impurity transport in the four stellarators, is addressed, providing the 
diffusion and convection coefficients for two selected impurities under pure ITG turbulence. 
In section \ref{sec:w7x_nscan_itg}, the linear stability of electrostatic modes in W7-X, when the density gradient of the main species is scanned at constant ion temperature gradient, is briefly examined and compared with LHD, {TJ-II} and NCSX. For that same scan, by means of nonlinear simulations with iron at trace concentration in W7-X only, question (ii) above, about the impurity transport in enhanced performance scenarios, is answered \reftwo{in section \ref{sec:w7x_nscan_gammaZ}}.
Finally, section \ref{sec:discussion} briefly summarizes the results.

\revone{All simulations presented in the paper, from sections \ref{sec:turbulence} to \ref{sec:w7x_nscan_gammaZ}, have been performed with the code \stella, which is briefly described before the first numerical results of the work are presented in section \ref{sec:turbulence}. All simulations have considered all species, namely, electrons, main ions and, when required, an impurity at trace concentration, as kinetic species. For discussion purposes a few linear simulations with adiabatic electrons have been included in section \ref{sec:w7x_nscan_itg}}.

\section{ITG stability and turbulence in W7-X, LHD, TJ-II and NCSX}
\label{sec:turbulence}

\begin{table}
	\begin{center}
		\small
		\begin{tabular}{ c | c c c c c c c }
			Device & $\{a, R_0\}$ [m]  & $N_{\mathrm{fp}}$ & $B_{r}$ [T] & $\iota$ & $\hat{s}$ &
			$\{N_x, N_y, N_z, N_{v_{\|}}, N_{\mu}\}^{\mathrm{nl}}$ & 
			$(k_y\rho_i)_{\mathrm{min}}^{\mathrm{nl}}$\\ [0.5 mm] \hline 
			NCSX   & \{0.323, 1.44\}    & 3    & 1.52    & 0.570   & -0.823 & $\left\{103,75,96,24,12\right\}$ &
			$0.067$\\ 		
			TJ-II  & \{0.193, 1.50\}    & 4     & 0.82    & -1.600 & -0.674 & $\left\{103,64,96,24,12\right\}$ &
			$0.100$\\  
			W7-X   & \{0.514, 5.51\}  & 5    & 2.64    & 0.886  & -0.150 & $\left\{76,109,96,24,12\right\}$ &
			$0.050$\\
			LHD    & \{0.636, 3.67\}    & 10   & 2.58    & -0.763 & -1.569 & $\left\{134,67,96,24,12\right\}$ &
			$0.067$
		\end{tabular}	
		\caption{For the stellarators W7-X, LHD, TJ-II and NCSX from left to right: effective minor radius ($a$) \revone{and major radius ($R_0$)}, number of field periods ($N_{\mathrm{fp}}$),volume averaged magnetic field ($B_r$, used as reference magnetic field value), rotational transform ($\iota$) and global magnetic shear ($\hat{s}$) at the simulated position $r/a=0.75$. Apart from these geometry related quantities, the resolution and minimum values $k_y$ set for the nonlinear simulations are provided in the two rightmost columns.}
		\label{tab:parameters}
	\end{center}
\end{table}

The $\delta f$ flux tube gyrokinetic code \stella\ \cite{Barnes_jcp_391_2019} has 
been employed to solve the Vlasov and quasi-neutrality equations. The spatial coordinates $\{x, y, z\}$ that the code uses for stellarator simulations are: the flux surface label $x=a\sqrt{s}$ (also denoted by $r$) with
$s=\psi_t/\psi_{t,\mathrm{LCFS}}$ the toroidal magnetic flux normalised 
to its value at the last closed flux surface; 
the magnetic field line label $y=a\sqrt{s_0}\alpha$, with $\alpha=\theta-\iota\zeta$, 
$\theta$ and $\zeta$ the poloidal and toroidal, respectively, PEST flux coordinates \cite{Grimm_jcp_49_1983}, $\iota$ the rotational transform and $s_0$ the 
value of the flux surface label around which the flux tube is centred; 
the parallel coordinate $z=\zeta$. The velocity coordinates
are the parallel velocity $v_{\|}$ \reftwo{and the magnetic moment $\mu=m_sv_{\bot}^2/2B$, where $m_s$ 
is the mass of species $s$, $v_{\bot}$ is the perpendicular velocity and $B$ is the magnetic field strength}. 
The mixed implicit-explicit algorithm employed by \stella~
to solve the gyrokinetic equation, in particular the implicit treatment of 
the parallel streaming and acceleration terms, allows to use larger time steps than what explicit methods permit, which is especially beneficial when kinetic species with disparate masses are considered.

\begin{figure}
	\centering
	\includegraphics[width=0.24\textwidth]{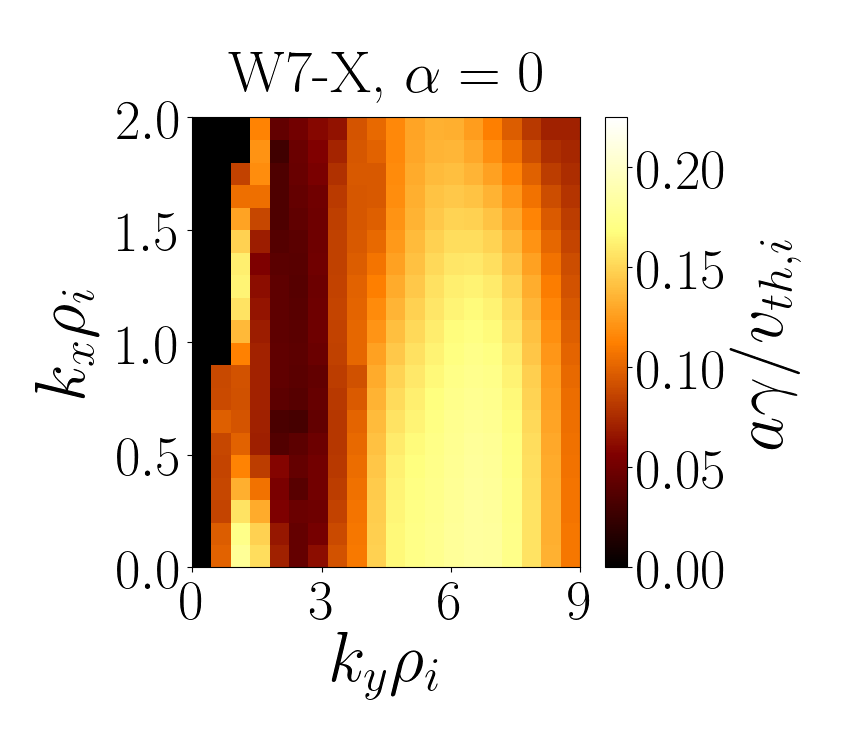}
	\includegraphics[width=0.24\textwidth]{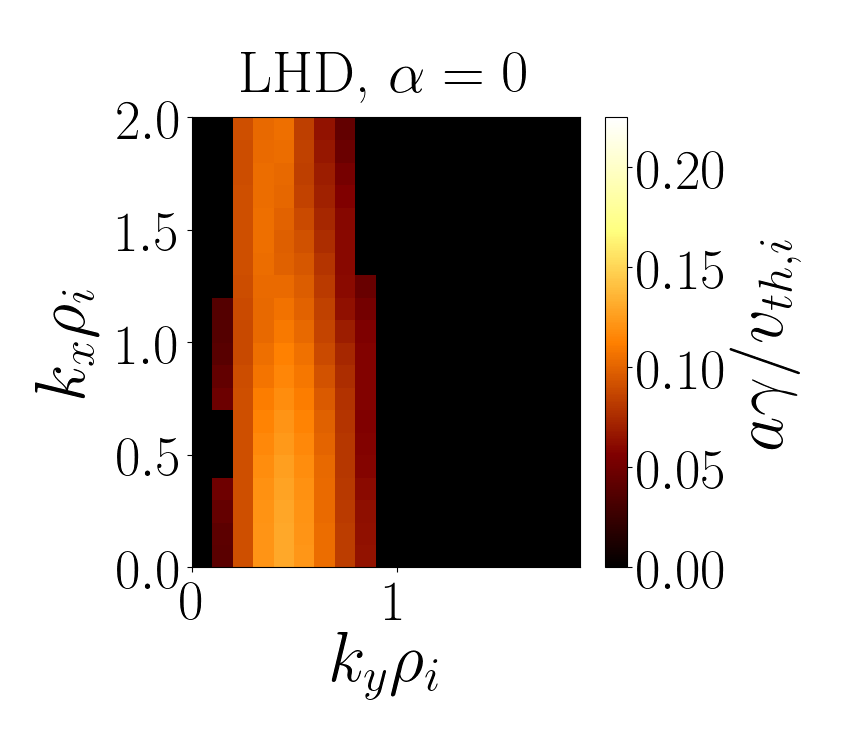}
	\includegraphics[width=0.24\textwidth]{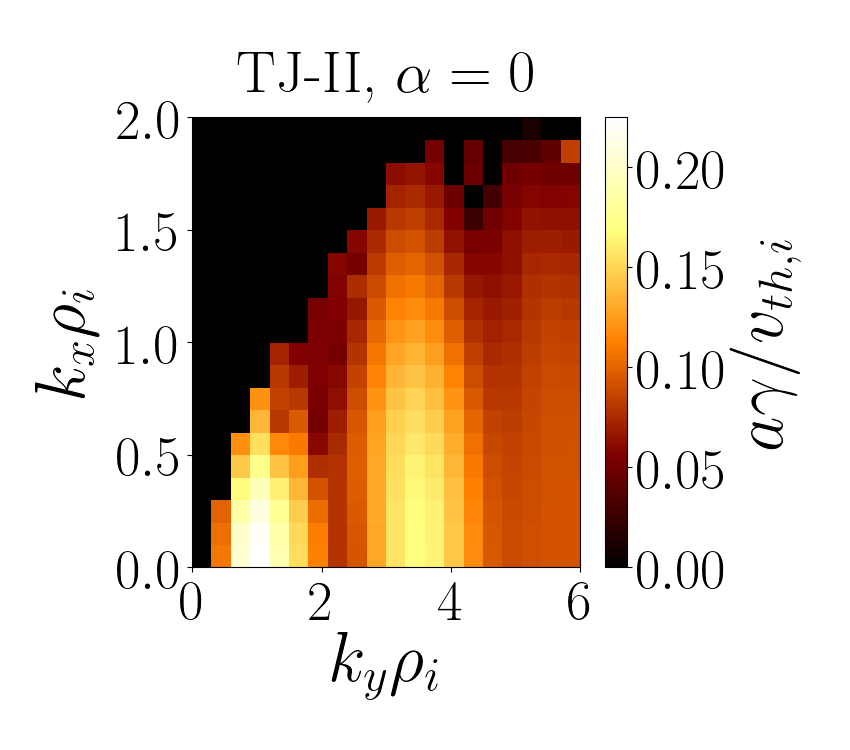}		\includegraphics[width=0.24\textwidth]{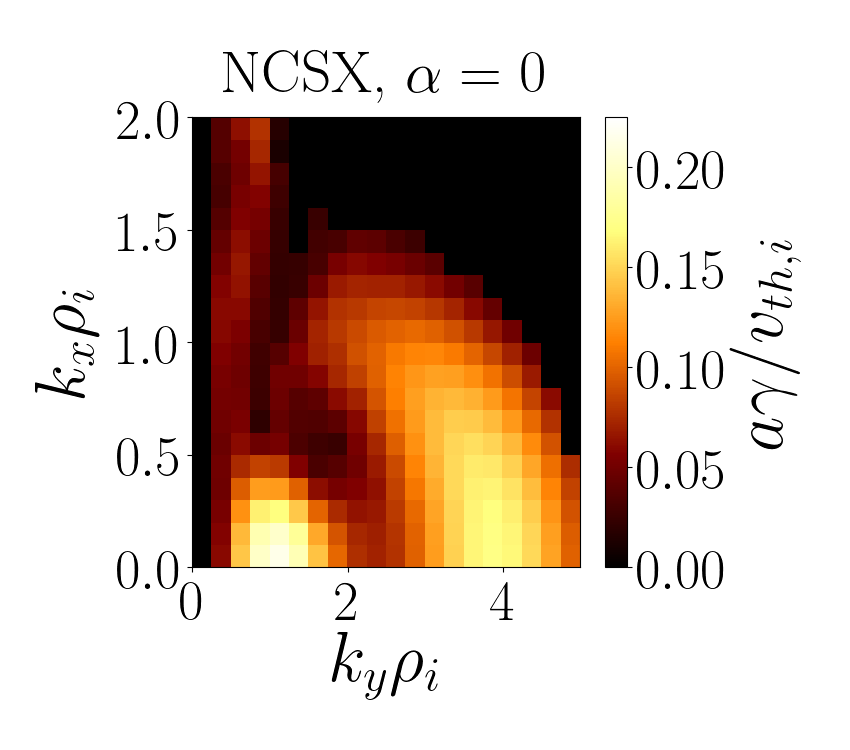}\\
	\includegraphics[width=0.24\textwidth]{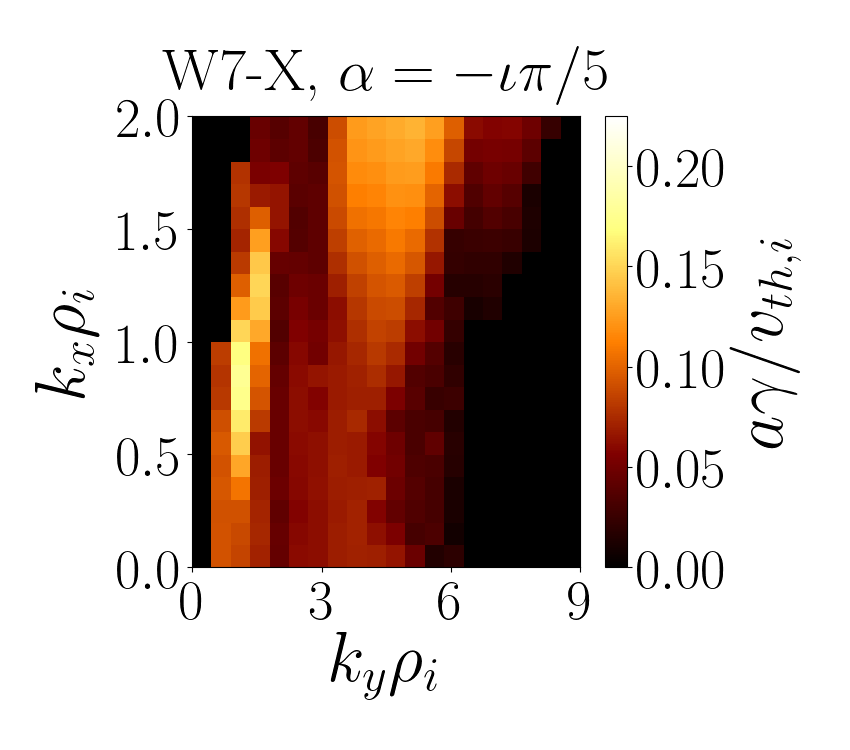}
	\includegraphics[width=0.24\textwidth]{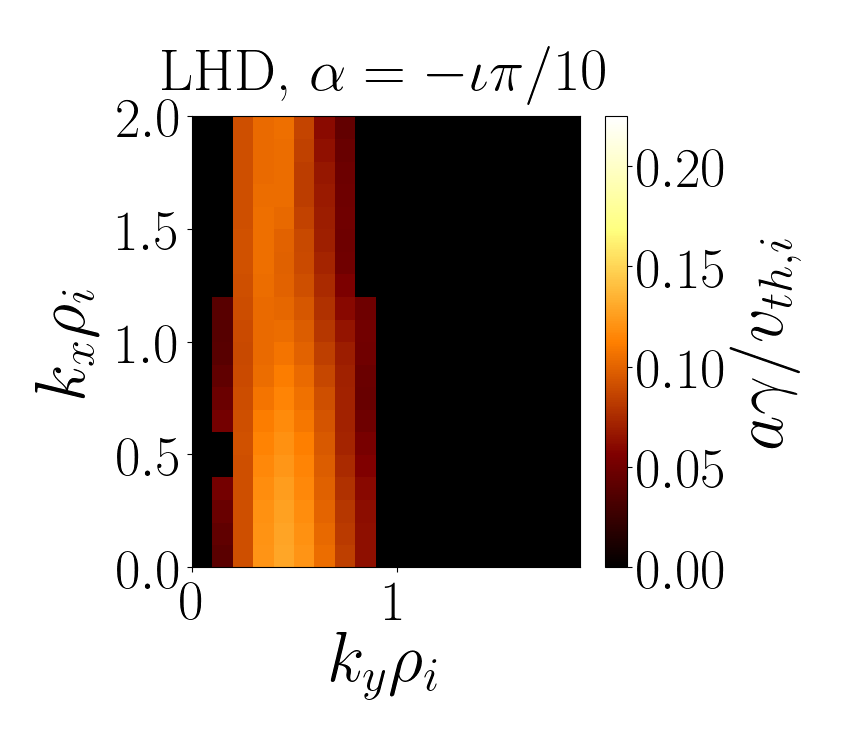}
	\includegraphics[width=0.24\textwidth]{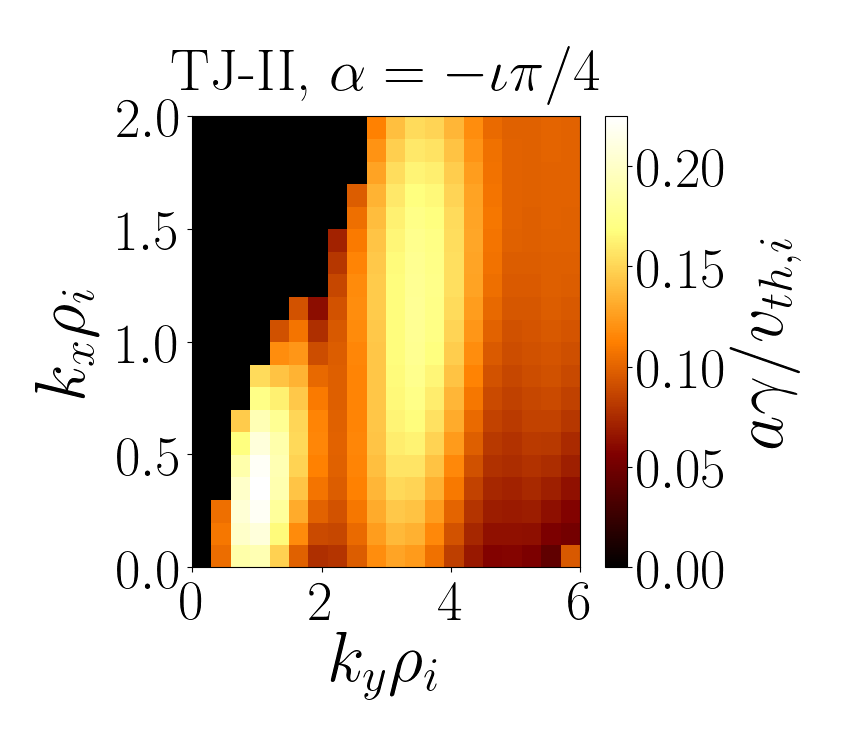}		\includegraphics[width=0.24\textwidth]{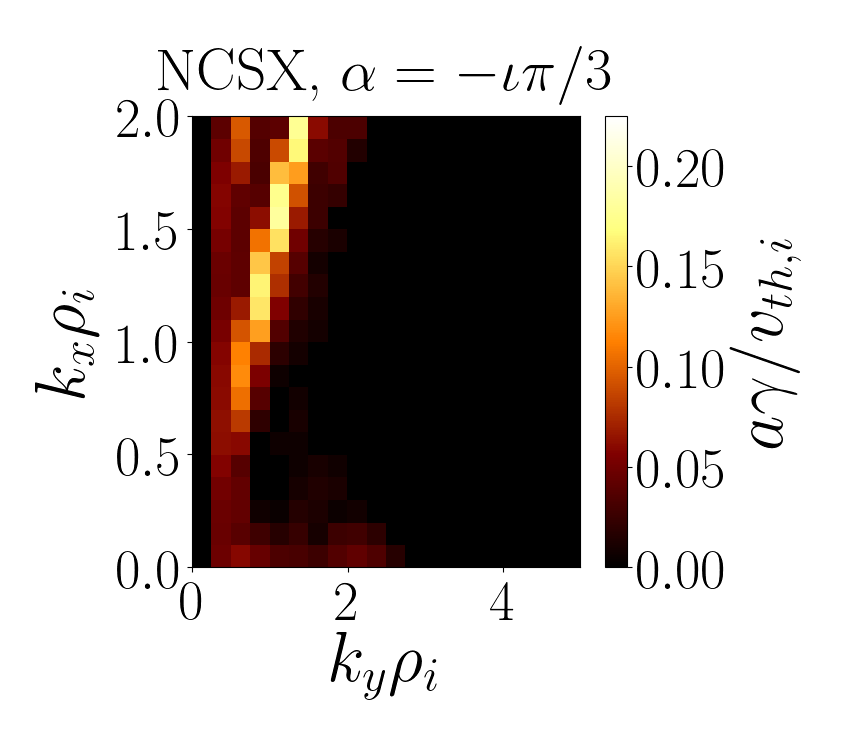}
	\caption{\blue{Normalised growth rate as a function of the radial and binormal wavenumbers, $k_x$ and $k_y$, respectively, for W7-X, LHD, TJ-II and NCSX (from left to right) for the flux tubes centered with respect the equatorial plane, $\theta=0$ and the toroidal planes $\zeta=0$ (top row) and $\zeta=\pi/N_{\text{fp}}$ (bottom row).}}
	\label{fig:kxky_linear}
\end{figure}

The input magnetic geometry for stellarator simulations is calculated by the ideal MHD code \texttt{VMEC} \cite{Hirshman_pop_28_1985}. For the comparison that follows, the standard magnetic configuration of W7-X (commonly referred to as EIM configuration) \cite{Geiger_ppcf_57_1_2015}, the inward shifted configuration of LHD \cite{Murakami2002}, the standard configuration of TJ-II \cite{Ascasibar2005} and a sample configuration of NCSX \cite{Nelson2003} have been considered, all with zero volume averaged plasma pressure. The main parameters of each configuration are listed in table \ref{tab:parameters}. The simulated radial position has been set to $r/a=0.75$. Finally, the gradients of the main species have been chosen so that a pure ITG instability is driven, i.e.~all gradients other than the ion temperature gradient have been set to zero:  \revone{$\left\{a/L_{T_i}, a/L_{T_e}, a/L_{n_e}, a/L_{n_i}\right\}=\left\{4.0, 0, 0, 0\right\}$}. For these conditions, prior to the nonlinear calculations performed to obtain the diffusion and the convection coefficients presented in the next section, a series of linear simulations have been carried out. For the phase space 
grid resolution of these simulations, the number of grid points along the parallel coordinate has been set to $N_z=384$, and for the velocity space grid $\{N_\mu,N_{v_{\|}}\}=\{12,24\}$ has been taken.
Defining the perpendicular wavevector as $\mathbf{k}_{\bot}=k_x\nabla x + k_y\nabla y$, linear simulations scanning $k_x$ and $k_y$ for each configuration have been performed for the flux tubes (extended them three turns poloidally) $\alpha=0$ and  $\alpha=-\iota \pi/N_{\mathrm{fp}}$, with $N_{\mathrm{fp}}$ the number of field periods of each configuration. The growth rate spectra resulting from these scans are represented in figure \ref{fig:kxky_linear} and show that, for W7-X, LHD, NCSX and TJ-II, $\alpha=0$ contains the most unstable mode at $k_x=0$. For the flux tube $\alpha=-\iota \pi/N_{\mathrm{fp}}$ in the cases of W7-X, NCSX and TJ-II the most unstable mode has $k_{x}\ne 0$, and its growth rate is, at most, comparable to the growth rate of the fastest growing mode found at $\alpha=0$. For LHD, the simulations show practically no differences in the localisation of the most unstable mode in Fourier space and its growth rate when moving from one flux tube to the other.
Thus, the flux tube $\alpha=0$ has been selected for all simulations that follow.

\begin{figure}
	\includegraphics[width=0.33\textwidth]{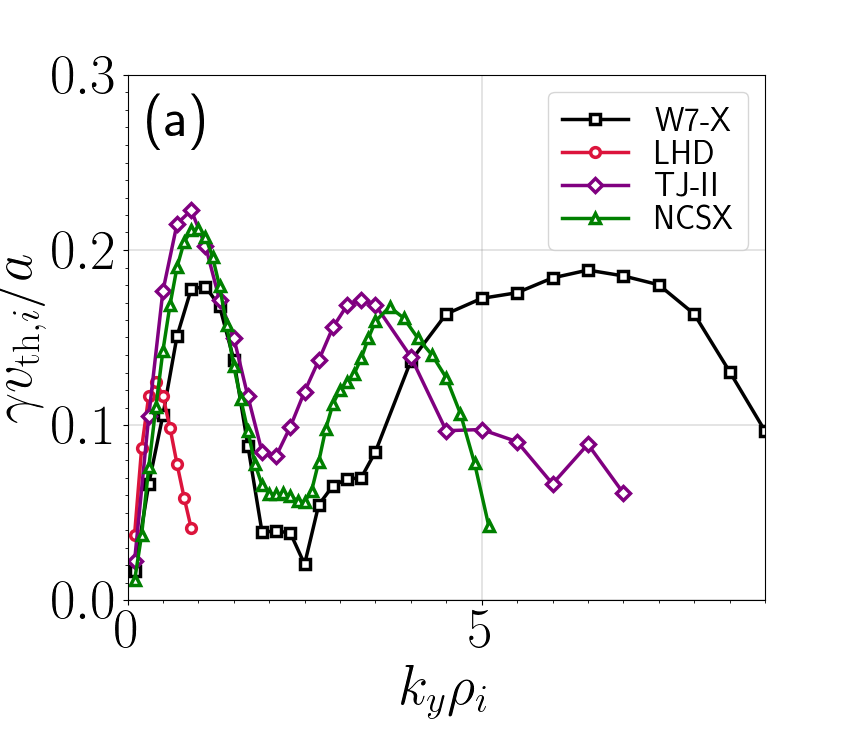}
	\includegraphics[width=0.33\textwidth]{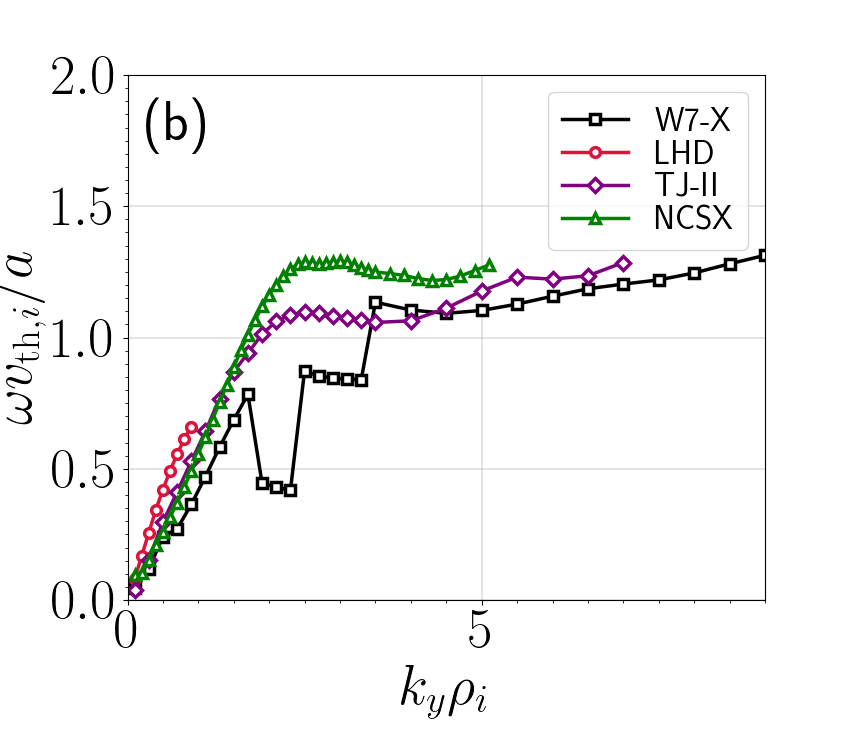}
	\includegraphics[width=0.33\textwidth]{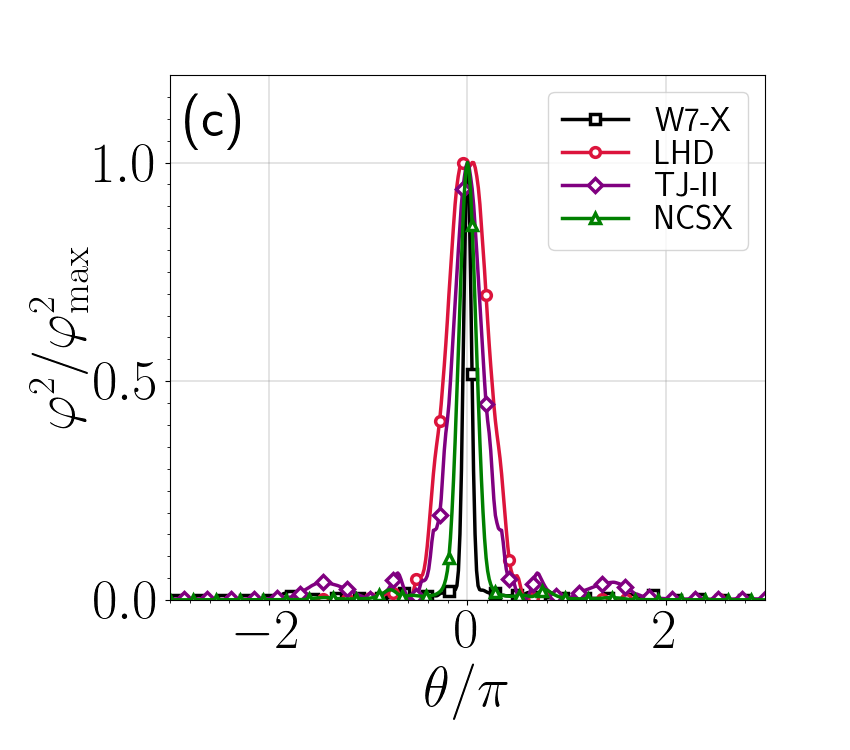}
	\caption{For the considered ITG conditions and the four stellarator devices: normalised growth rate \blue{(a)} and frequency \blue{(b)} as a function of the binormal wavenumber $k_y$ for $k_x=0$; parallel mode structure for the fastest growing mode for each device \blue{(c)}.}
	\label{fig:linear}	
\end{figure}

Particularizing then for the flux tube $\alpha=0$ and $k_x=0$, the results from the linear simulations are represented in figure \ref{fig:linear} and, regarding the stability of the ITG case considered, they exhibit significant differences across the four stellarators. Looking at the panel \blue{(a)}, one can observe that unstable modes are found along a broad $k_y$ range for TJ-II, NCSX and, particularly, for W7-X. The growth rate, which for the three devices just mentioned exhibits two local maxima, has comparable sizes for the most unstable modes, just slightly larger for TJ-II than for NCSX, for which, in turn, the growth rate is also appreciably larger than for W7-X.
In contrast, the same conditions produce in LHD a significantly narrower $k_y$ spectrum. In addition, the growth rate of the most unstable mode in LHD is approximately half of that of the most unstable modes in the other three configurations, which represents a remarkable difference. On the other hand, the parallel structure of the most unstable mode, depicted in figure \ref{fig:linear} \blue{(c)} is rather localised around the centre of the flux tube in the four cases, but especially in NCSX and W7-X. Similarly, the frequency, shown in figure \ref{fig:linear} \blue{(b)}, does not display any particularly strong difference in its shape, with the exception of the discontinuous pattern for W7-X, indicative of successive changes of dominant mode along the simulated $k_y$ interval \refone{(for a discussion about the changes in the parallel structure of ITG modes in W7-X, see e.g.~\cite{Zocco2018})}. 

\begin{figure}
	\hspace{0.15cm}\includegraphics[width=0.24\textwidth]{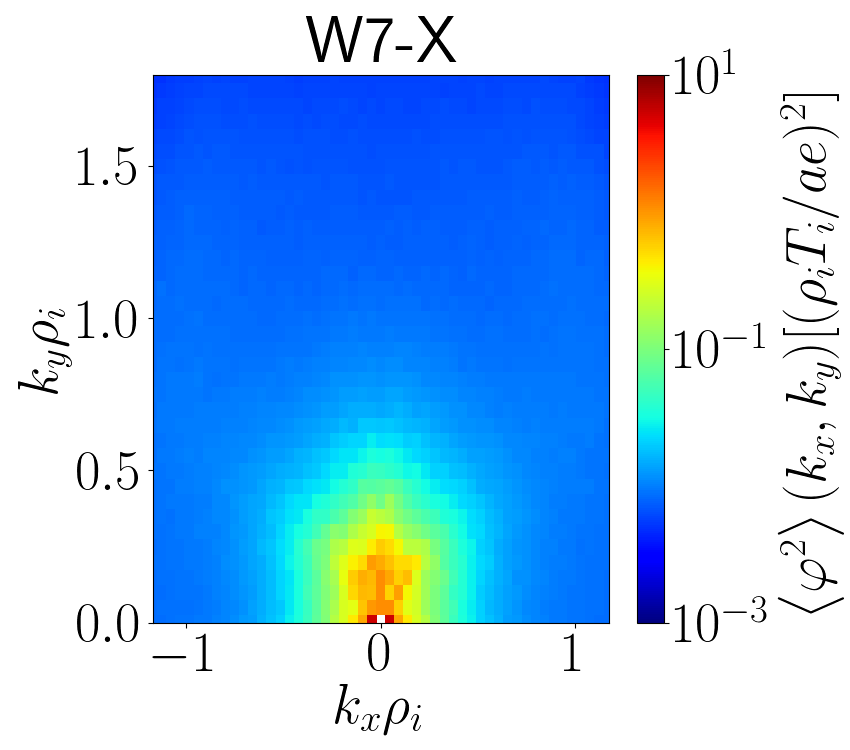}
	\includegraphics[width=0.242\textwidth]{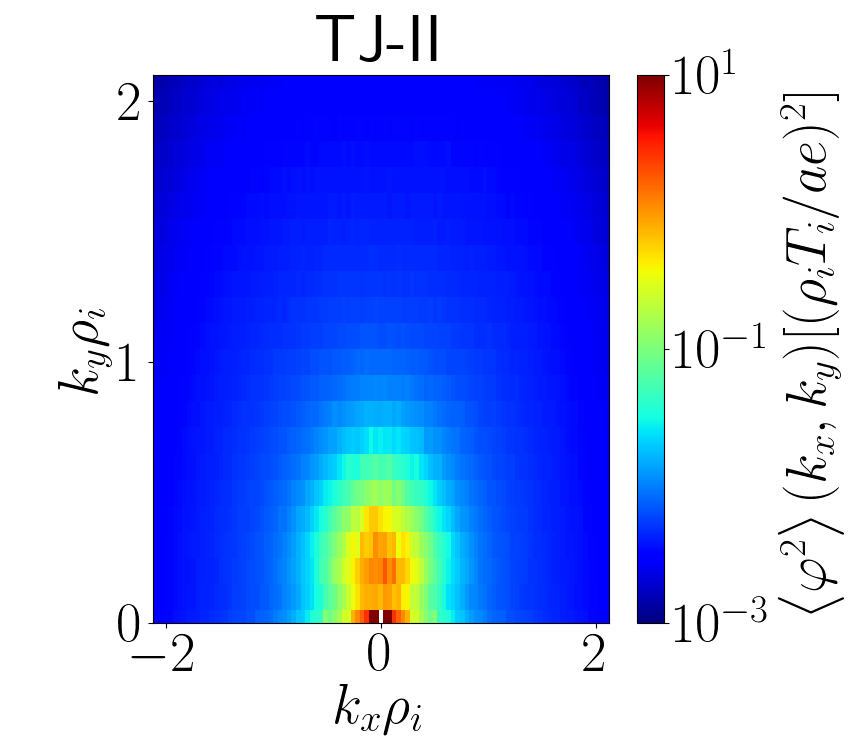}
	\includegraphics[width=0.242\textwidth]{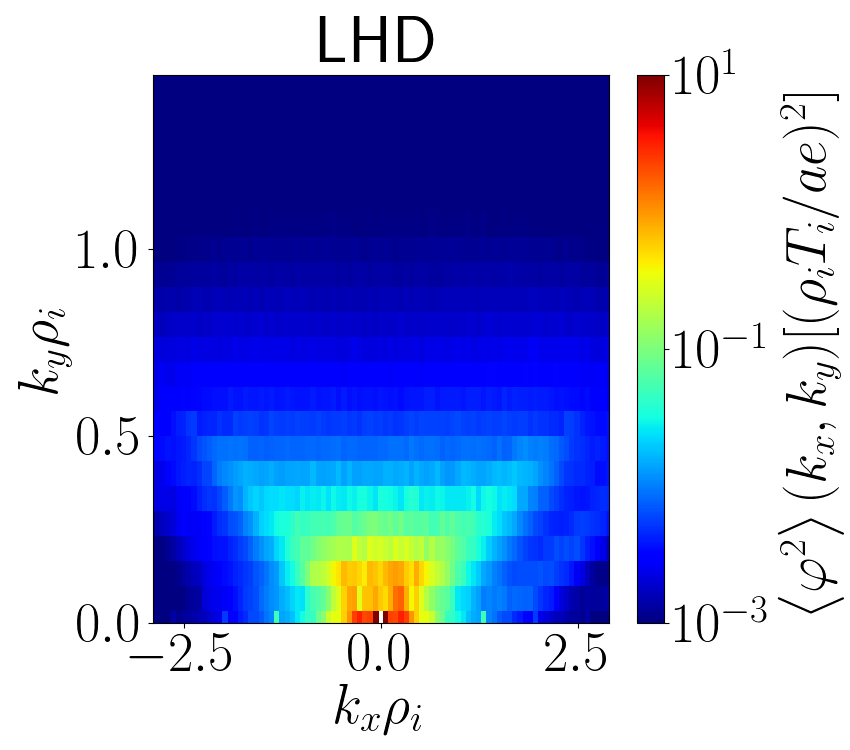}
	\includegraphics[width=0.242\textwidth]{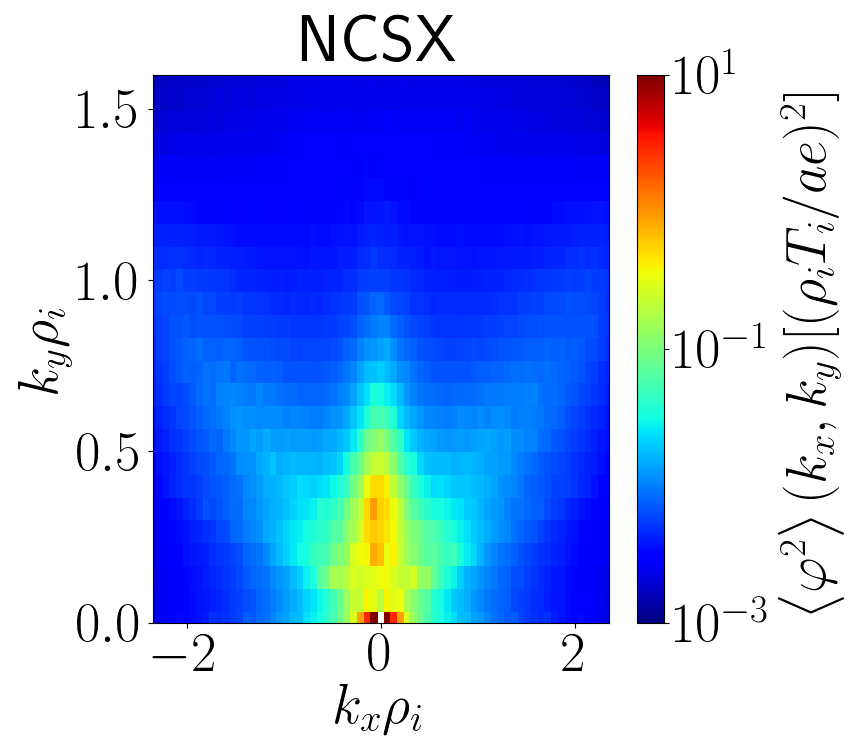}\\
	\includegraphics[width=0.242\textwidth]{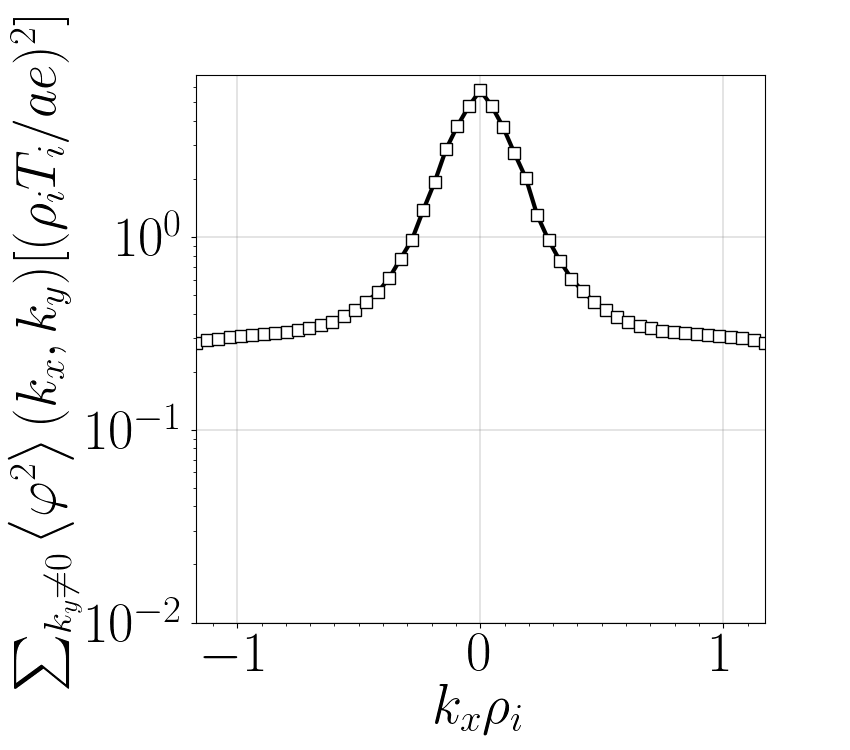}
	\includegraphics[width=0.24\textwidth]{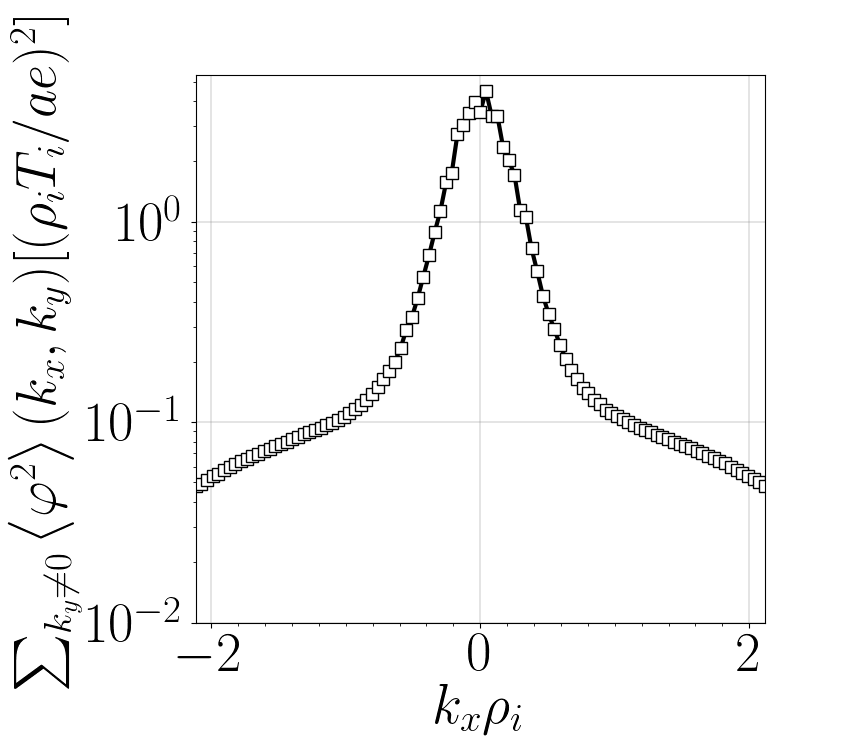}
	\includegraphics[width=0.24\textwidth]{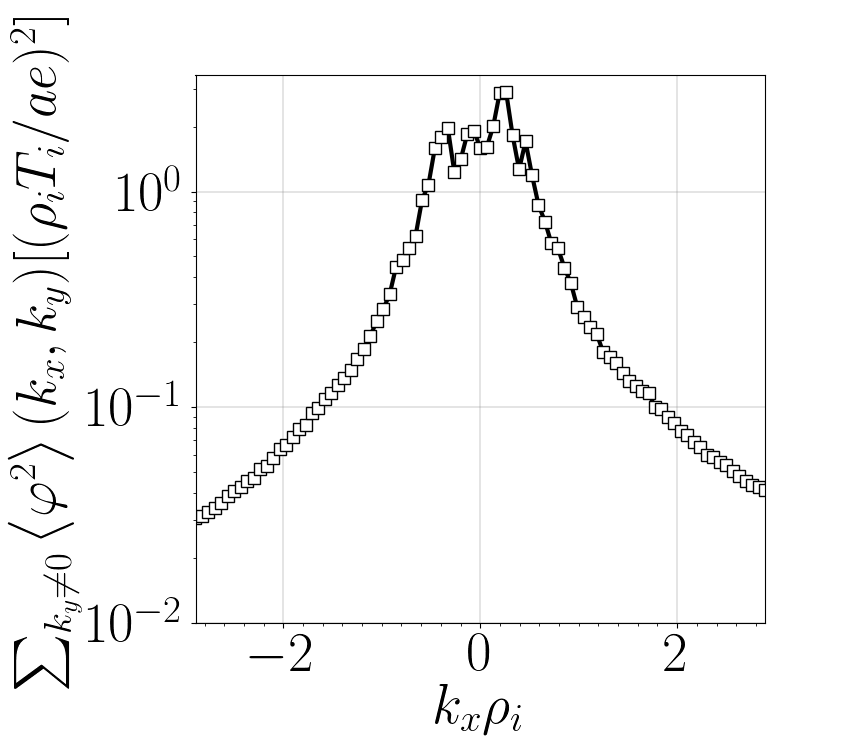}
	\includegraphics[width=0.24\textwidth]{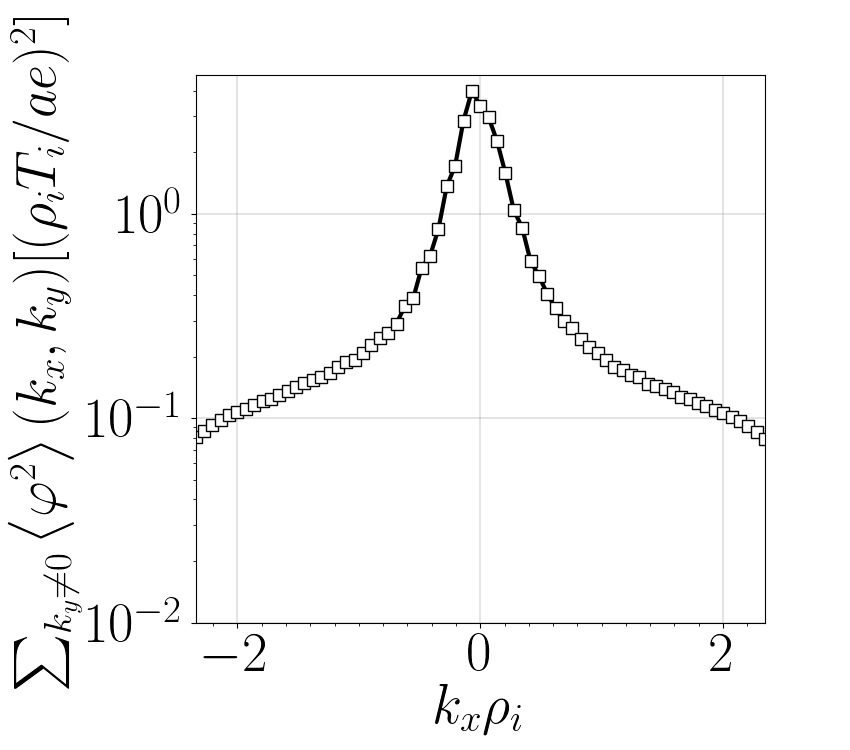}\\
	\includegraphics[width=0.24\textwidth]{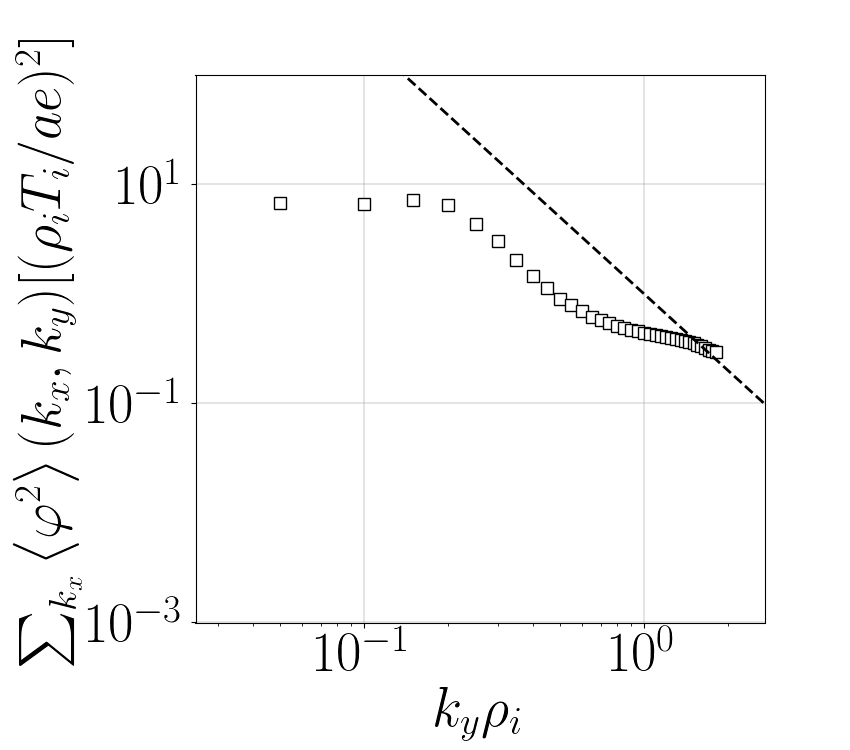}
	\includegraphics[width=0.24\textwidth]{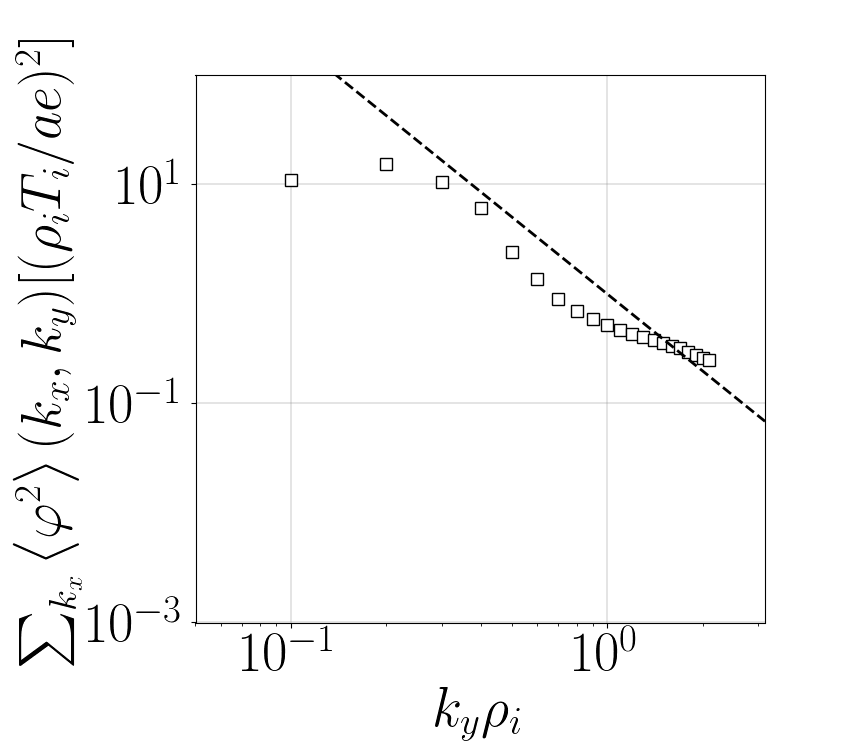}
	\includegraphics[width=0.24\textwidth]{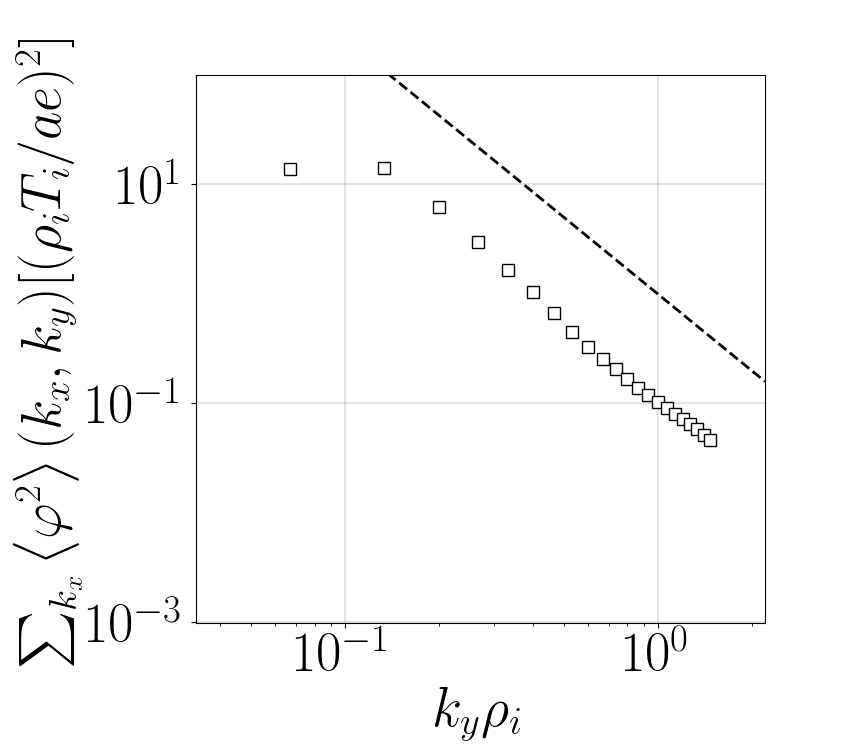}
	\includegraphics[width=0.24\textwidth]{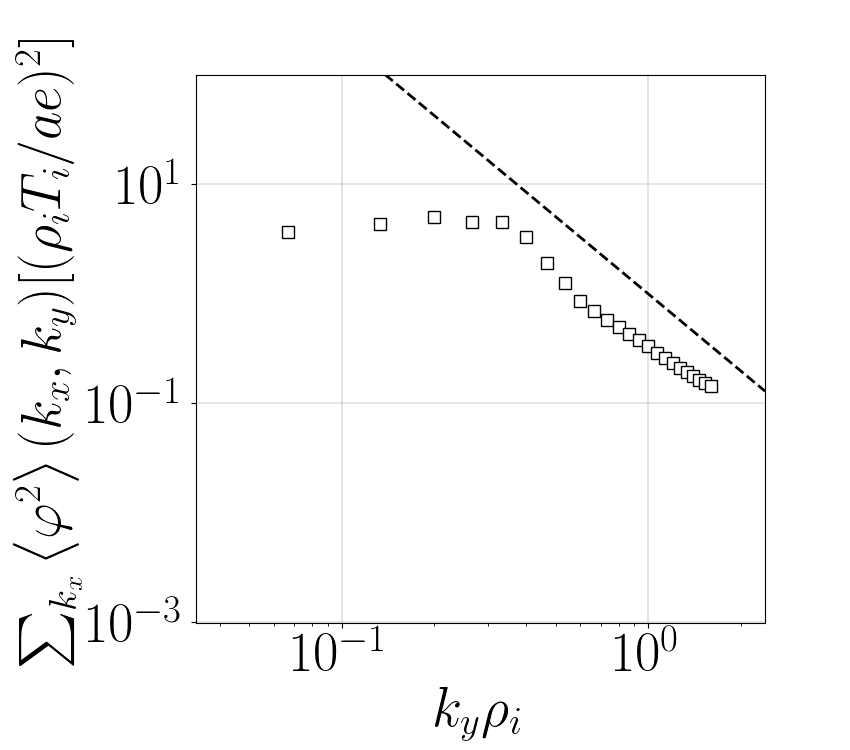}
	\caption{From top to bottom: spectrum of the field line averaged value of the square of the turbulent electrostatic potential, $\left<\varphi^2\right>(k_x,k_y)$; $k_x$ spectrum of $\left<\varphi^2\right>$ neglecting the zonal components, obtained as $\sum_{k_y\ne 0}\left<\varphi^2\right>(k_x,k_y)$; $k_y$ spectrum of $\left<\varphi^2\right>$, obtained as $\sum_{k_x}\left<\varphi^2\right>(k_x,k_y)$, and power law $(k_y\rho_i)^{-7/3}$ indicated with a dashed line. The results are displayed, from left to right, for W7-X, TJ-II, LHD and NCSX.}
	\label{fig:kxky_nonlinear}	
\end{figure}

Nonlinearly, the simulations have been performed for the flux tube $\alpha=0$, extended along one poloidal turn. While for all stellarators the resolution in the velocity space grid and parallel coordinate has been the same,  $\{N_\mu,N_{v_{\|}}\}=\{12,24\}$ and $N_z=96$, the width and the number of divisions along the $x$ and $y$ directions ($N_x$ and $N_y$, respectively) have been set differently for each of the four devices. \revone{The resolution chosen for each case, to ensure the fluxes are well converged and the electrostatic potential is well bounded within the box in Fourier space, is provided in table \ref{tab:parameters}}. In figure \ref{fig:kxky_nonlinear} (top), the spectrum of the field line averaged value of the square of the turbulent potential, $\left<\varphi^2\right>(k_x,k_y)$, is shown. In the four figures, it can be examined that the amplitude of the modes extends through two orders of magnitude, approximately. Regarding the shape of the spectra, it can be appreciated that the largest components corresponds to zonal modes ($k_y=0$) with low $k_x$ values. On the other hand, the non-zonal components of TJ-II and W7-X exhibit a strong concentration in a relatively narrow region of the considered $k_x$ interval, while LHD and NCSX show a spectrum that broadens as $k_y$ increases. Looking at the central and bottom row of figure \revone{\ref{fig:kxky_nonlinear}}, one can inspect the spectra in $k_x$ and $k_y$, respectively. Again, although the $k_x$-spectra are rather similar for W7-X, NCSX and TJ-II, the spectrum of LHD features slight deviations with respect to the shape of the other three devices, being less peaked around $k_x=0$ and decaying faster at larger $|k_x|$ values. Regarding the $k_y$ spectra, a slight deviation from the power law $(k_y\rho_i)^{-7/3}$ \cite{Barnes_prl_2011} \revone{can be appreciated} for W7-X\revtwo{, which does not affect the fluxes and can be removed by increasing the resolution \cite{Regana_JPP_2021}}, while TJ-II, LHD and NCSX follow it more tightly.

\section{Transport of impurities in W7-X, LHD, TJ-II and NCSX driven by ITG turbulence}
\label{sec:transport} 

\begin{figure}
	\includegraphics[width=0.5\textwidth]{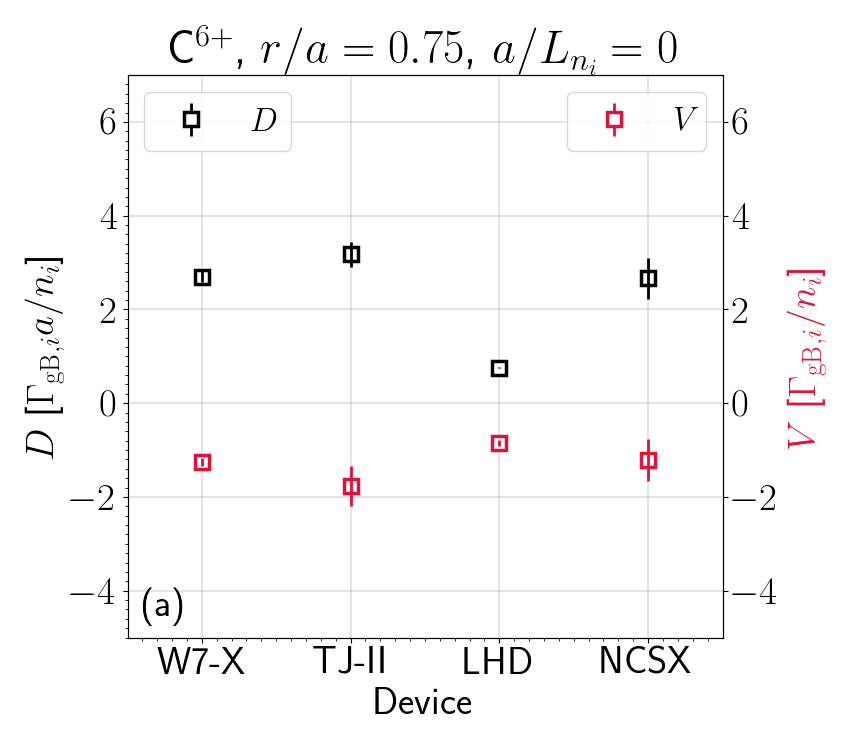}
	\includegraphics[width=0.5\textwidth]{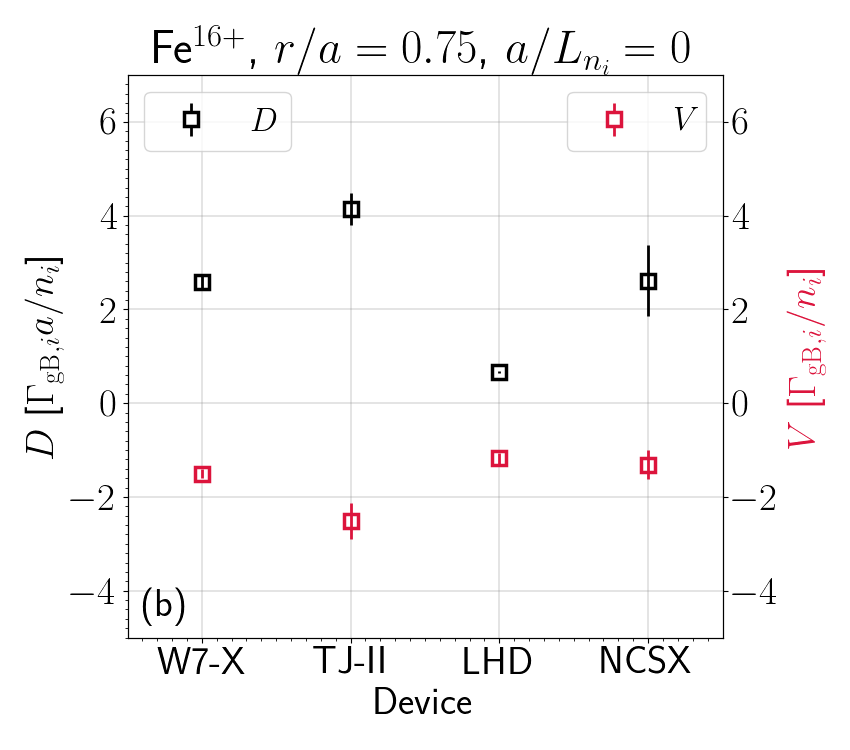}
	\caption{For $C^{6+}$ (left) and Fe$^{16+}$ (right), ITG-driven diffusion coefficient ($D$) and convection velocity ($V$) in normalised units at the position $r/a=0.75$ for W7-X, TJ-II, LHD and NCSX, with $\Gamma_{\mathrm{gB,i}}$ the gyro-Bohm particle flux for the main ions.}
	\label{fig:D_and_V}
\end{figure}

The most frequent situation in the W7-X plasmas during the first campaigns, mostly subject to strong ECRH, has resulted in a  neoclassical electric field (and convection connected to it) typically positive (electron root) at the inner core and negative (ion root) in the outer mid plasma radius \cite{Pablant_pop_25_2018}. Consequently, severe accumulation of highly charged impurities in the plasma core may have been prevented so far, in part, due to the core electron root confinement  (CERC) conditions \cite{Klinger_nf_59_112004_2019}. Nonetheless, in the outer mid region of those plasmas, which are under ion-root conditions and where inward neoclassical convection of impurities would take place, peaked impurity density profile formation has not been observed. The turbulent diffusion coefficient, inferred from the experimental measurements \cite{Geiger_NF_59_046009_2019} and confirmed numerically \cite{Regana_JPP_2021} in the edge region, seems consistently large to guarantee that the turbulent or neoclassical convective sources, when inwards, cannot yield strongly peaked impurity profiles.
The aim of the present section is to quantitatively assess whether the transport coefficients (in particular the diffusion coefficient) in W7-X are, in comparison with the other three stellarators presented in section \ref{sec:turbulence}, actually larger. For this purpose, the nonlinear simulations have included, apart from the main ions and electrons, an impurity species at \refone{a trace concentration of $n_{Z}/n_{i}=10^{-12}$, with $n_Z$ and $n_i$ the background densities of the impurity and the main ions, respectively}. The three species present in the problem have been treated kinetically. Finally, fully ionised carbon and iron with $Z=16$ have been considered for the results presented in this section.  Being the normalised impurity particle flux expressed as $\Gamma_Z/n_Z=-D\text{d}\ln(n_Z)/\text{d}r+V$, the temperature gradient of the impurity species has been set to the same value as the ion temperature gradient, \blue{$a/L_{T_i}=a/L_{T_Z}=4.0$}, while two different values for the impurity density gradient, $a/L_{n_Z}$, have been necessary in order to obtain $D$ and $V$ for each impurity and magnetic configuration. \blue{As in the previous section, vanishing density gradients for the main ions and electrons, \revone{$a/L_n=0$}, and electron temperature gradient, $a/L_{T_e}=0$, have been \revone{considered}, in order to isolate the effect from a pure ITG mode.} Figure \ref{fig:D_and_V} \blue{(a)} summarises the results for C$^{6+}$ and figure \ref{fig:D_and_V} \blue{(b)} does it for Fe$^{16+}$. The two figures represent in normalised units the diffusion coefficient referred to the left y-axis and the convection velocity to the right y-axis\footnote{\refone{It is important to note the employment of the main ion gyro-Bohm particle flux $\Gamma_{gB,i}$, and the minor radius of the device, $a$, to normalise the particle flux of the simulated species and the length scales, respectively. As $a$ is different across the four devices as well as $\Gamma_{gB,i}=n_i v_{\mathrm{th,i}}\rho_i^2 a^{-2}$, differences in the transport coefficient between them can arise when expressing the results in SI units, considering the characteristic plasma parameters, size and magnetic field strength of each of them. Hence, for cross device comparisons to be meaningful, this normalisation, which accounts for the ordering of the different quantities involved in the problem, must be applied. Finally, in the expression for $\Gamma_{gB,i}$ above, for the main ion species, $v_{\mathrm{th,i}}=\sqrt{2T_i/m_i}$ is the thermal speed, $m_i$ is the mass, $T_i$ is the temperature, $\rho_i=v_\mathrm{th,i}/\Omega_i$ is the thermal Larmor radius, $\Omega_i=e B_{\mathrm{r}}/m_i$ is the gyro-frequency, $e$ is the unit charge and $B_{\mathrm{r}}$ is the reference magnetic field (indicated in table \ref{tab:parameters}).}}. Looking at the two plots, it can be observed that the differences due to the species are minor, and that the relative size of $D$ or $V$ across the four configurations is roughly the same for carbon and for iron. A common feature in the four devices is that ITG turbulence drives net inward impurity transport and, consequently, would yield negative impurity density gradients in equilibrium.
Interestingly, although the convection coefficient clearly shows larger values for TJ-II than for W7-X and NCSX, the absolute value of the convection velocity is also larger, apparently, in a comparable degree. Indeed, if we compute the turbulent peaking factor for (W7-X, TJ-II, LHD, NCSX) it results to be $(V/D)=\blue{-a^{-1}(0.47, \revtwo{0.56}, 1.12, 0.44)}$  for C$^{6+}$ and $(V/D)=\blue{-a^{-1}(0.59, \revtwo{0.44}, 1.77, 0.50)}$ for Fe$^{16+}$. For each of the two species, the numbers are surprisingly close to each other for the three stellarators other than LHD. In LHD, the relative size of $V$ and $D$ is such that ITGs would drive stronger impurity density gradients by a factor between 2 and 3,  approximately. This feature can be added up to the characteristics pointed out in the previous section on the essentially different character of the ITG turbulence and stability for LHD. Finally, as it can be concluded from this brief analysis, the large diffusion coefficient is not specific of W7-X, when compared with NCSX and TJ-II. In any case, it is not the size of the diffusion or convection coefficient separately what matters if, as it has been numerically shown, the absolute value of both transport coefficients enhances or reduces coherently.

\section{The role of the density gradient of the main species in mixed ITG/TEM stability in W7-X, LHD, TJ-II and NCSX }
\label{sec:w7x_nscan_itg}

\begin{figure}
	\includegraphics[width=0.5\textwidth]{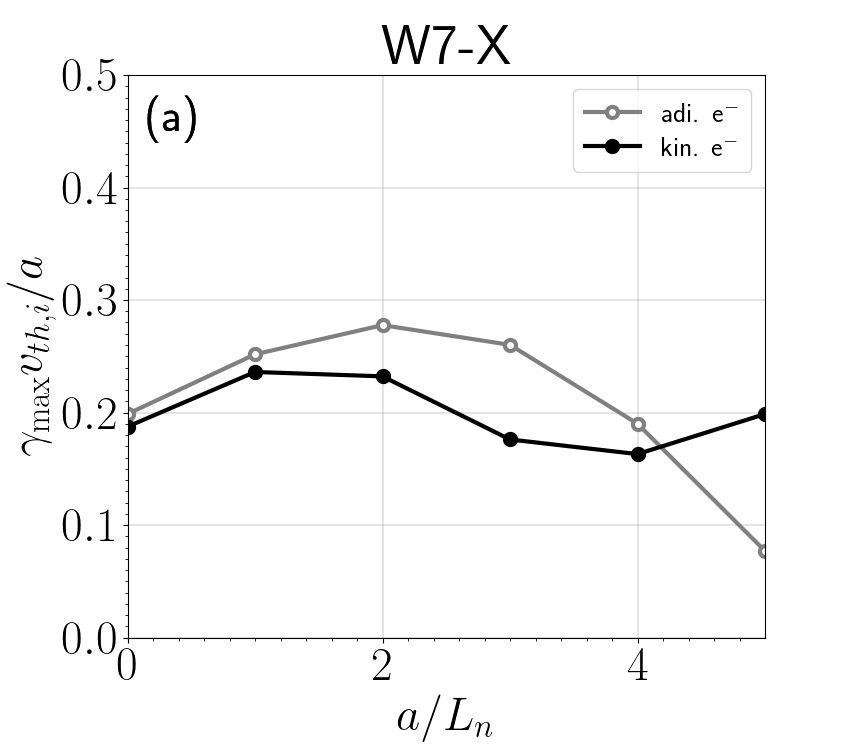}
	\includegraphics[width=0.5\textwidth]{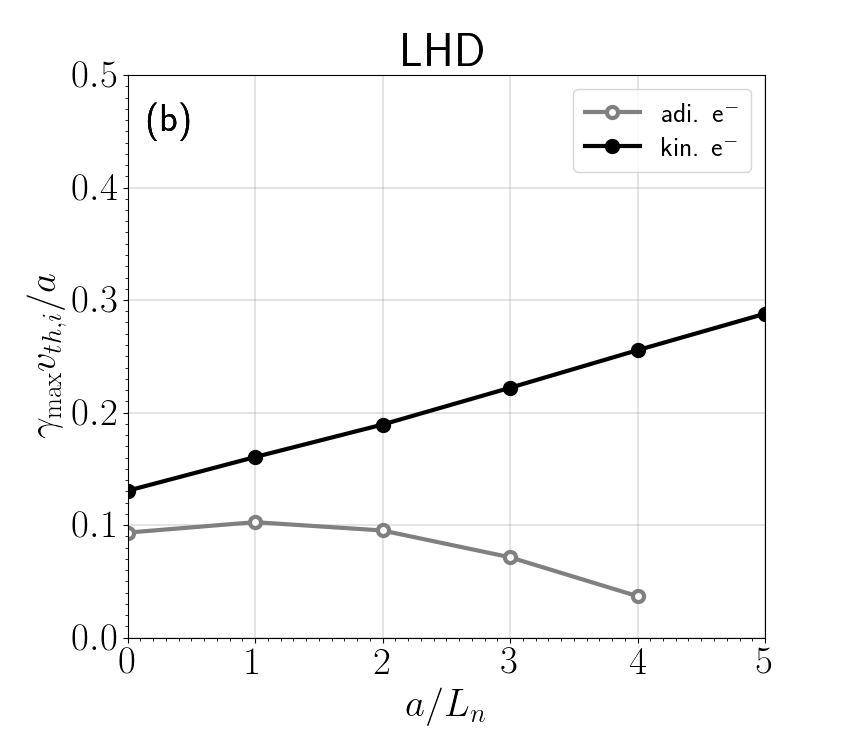}\\
	\includegraphics[width=0.5\textwidth]{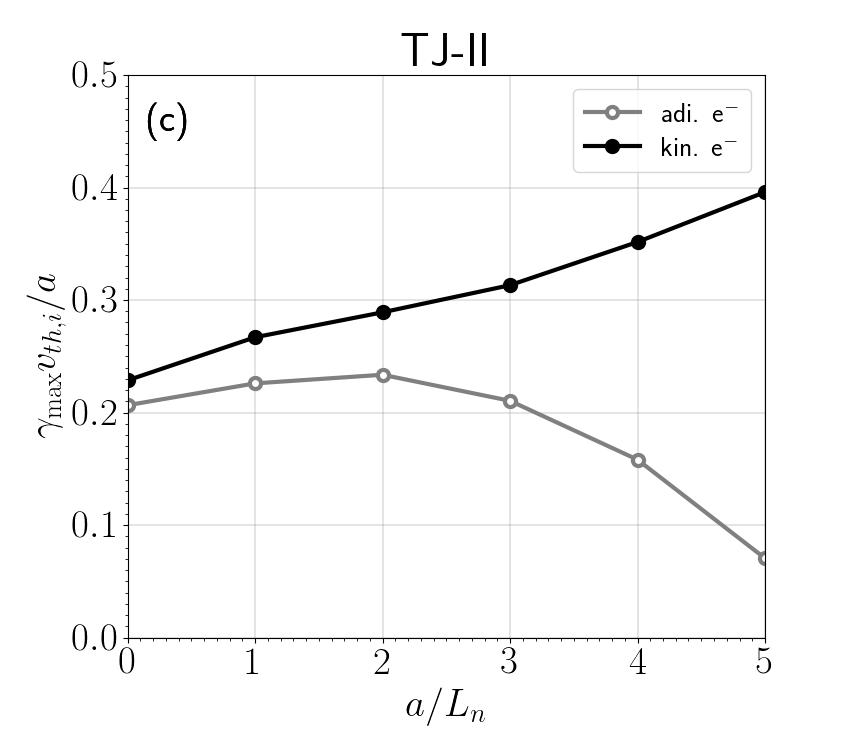}
	\includegraphics[width=0.5\textwidth]{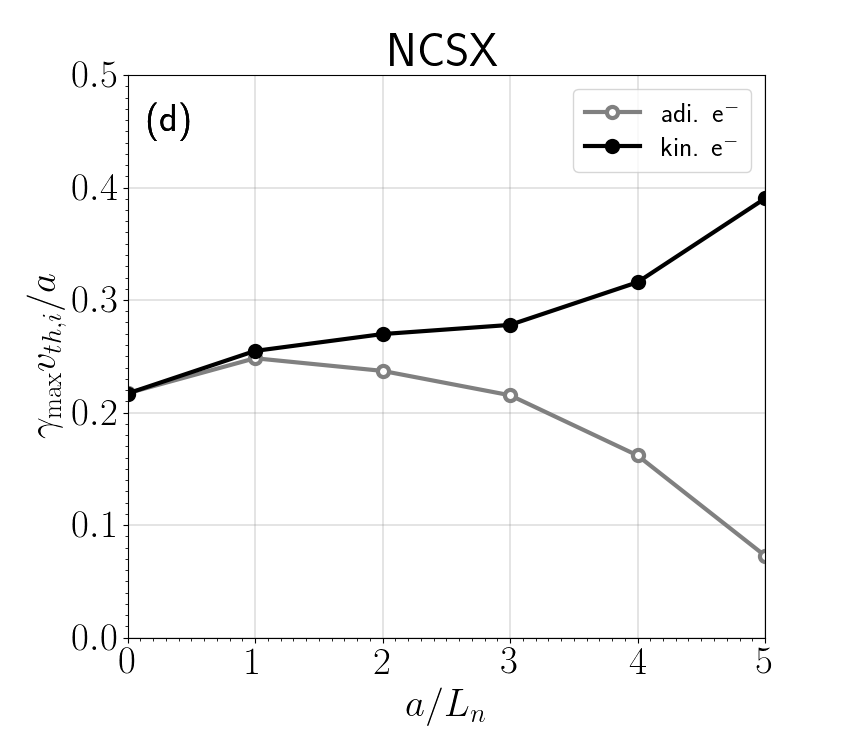}
	\caption{\blue{Maximum growth rate as a function of the normalised density gradient of the main ions and electrons, $a/L_n$, for an ion temperature gradient $a/L_{T_i}=4.0$, for the flux tube $\alpha=0$, taking the radial wavenumber $k_x=0$ and considering W7-X (a), LHD (b), TJ-II (c) and NCSX (d), including kinetic electrons (filled symbols) and assuming adiabatic electrons (open symbols).}}
	\label{fig:spectra}	
\end{figure}

\blue{During the first experimental campaigns, W7-X has accessed scenarios with record
values of triple product, stored energy and, for the modest installed (primarily ECRH) power, high core ion temperature \cite{Wolf_nf_57_102020_2017,Pedersen_ppcf_2018}. These so-called high performance scenarios are achieved by means of cryogenic pellet injections, which increase\refone{, practically at all radii above $r/a\sim 0.2$,} the density gradient of the main plasma species and lead to a reduction of the turbulent heat transport. In these situation, where power balance analyses \cite{Bozhenkov_nf_2020} and turbulent fluctuation measurements \cite{Estrada_nf_2021,Stechow_submitted_2021} confirm the reduced turbulence levels, the neoclassical optimisation targeted in the design of W7-X has been demonstrated \cite{Beidler_nature_submitted_2021}. On the other hand, during the enhanced confinement phase, longer impurity confinement time \cite{Stechow_submitted_2021} and eventually large impurity density gradients \cite{Langenberg_IAEA_2021} are measured.}

Linear gyrokinetic simulations show that W7-X enjoys lower growth rates of the electrostatic gyrokinetic instabilities as the density gradient, $a/L_n$, increases and approaches the value of the main ion temperature gradient $a/L_{T_i}$ \cite{Alcuson_ppcf_62_2020}. 
This feature has been postulated as a possible reason underlying the enhanced performance of those plasmas. Nonlinear 
simulations have supported this hypothesis as the increase of ion temperature and density gradients yields a reduction --significantly stronger than that experienced by the growth rate-- of the estimated ion heat flux \cite{Xanthopoulos_PRL_2020}. Regarding how the transport of impurities react to those reduced turbulence conditions remains unexplored, and the present section is devoted to shed light on that question.

\begin{figure}
	\begin{center}
	\includegraphics[width=0.5\textwidth]{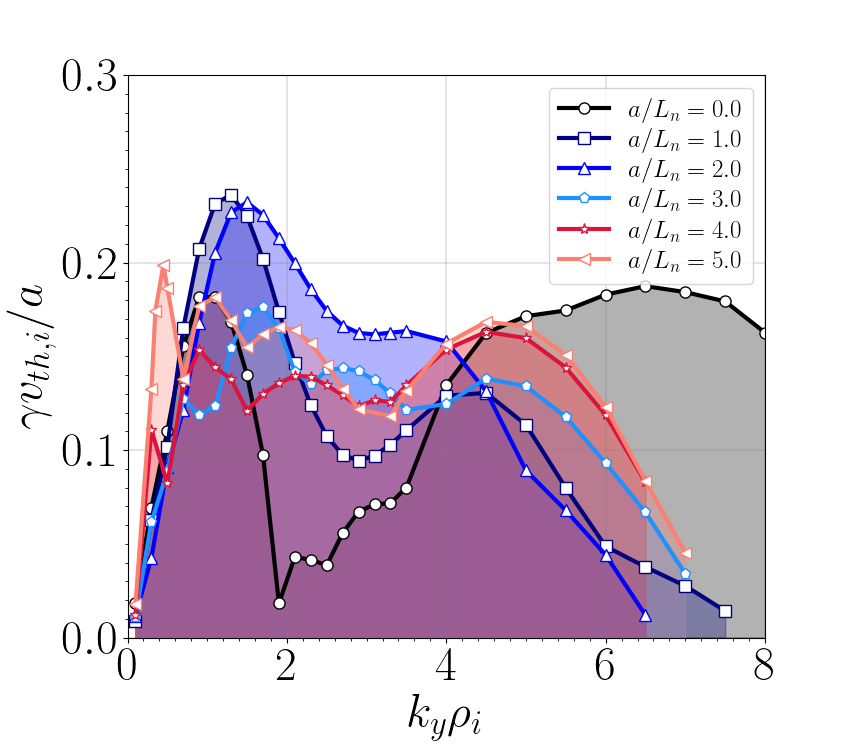}
	\caption{\blue{Growth rate $k_y$-spectrum at $r/a=0.75$, for the bean flux tube of W7-X and for the cases with $a/L_{T_i}=4.0$ and different value of normalised density gradient of the main species, $a/L_n=\left\{0, 1, 2, 3, 4, 5\right\}$.}}
	\label{fig:w7x_gamma_spectra}
	\end{center}
\end{figure}

For an ion temperature gradient of $a/L_{T_i}=4.0$, figures ~\ref{fig:spectra} (a)-(d) represent the maximum growth rate as a function of the density gradient $a/L_{n}$ for each of the four stellarators analysed. Comparing the growth rate with kinetic and adiabatic electrons, the distinctive character of W7-X electrostatic instabilities is clearly manifested. In the first place, the presence of kinetic electrons contribute to decrease the growth rate of the most unstable mode with respect to the case with adiabatic electrons. Only for $a/L_n=5$ the fastest growing mode is more unstable when including kinetic electrons than when they are assumed adiabatic. In the other three stellarators, the presence of kinetic electrons contribute to increase the growth rate of the most unstable mode to the point that it monotonically increases with $a/L_n$. In summary, in contrast to LHD, NCSX or TJ-II, ITG modes undergo a stabilisation in W7-X driven by the density gradient before this can noticeably destabilize TEMs. 

\section{Transport of impurities in reduced ITG turbulence scenarios in W7-X}
\label{sec:w7x_nscan_gammaZ}

To what extent a reduction in the maximum growth rate (of approximately 30\% from its largest value at $a/L_n=1.0$ to its minimum at $a/L_n=4$, seen in fig.~\ref{fig:spectra} (a)) can be reflected in other quantities, like the diffusion coefficients of the impurities, is difficult to extrapolate as the subtleties are numerous. Looking at the $k_y$-spectra of the growth rate for W7-X, represented in figure \ref{fig:w7x_gamma_spectra}, there are significant changes in their shape, which can exhibit a single maximum (for $a/L_n=2$), having several local maxima with comparable size (like $a/L_n=0$ or  $a/L_n=5.0$) or spread up to rather large $k_y$ values ($a/L_n=0$). Not surprisingly, conclusions based on linear results and mixing length arguments are not necessarily reflected in nonlinear simulations and can even be contradictory. \revone{For instance, in ref.~\cite{Regana_JPP_2021} the quasilinear calculations capture only qualitatively some of the results obtained for the impurity transport coefficients, like their relative size, but others, like the convection driven by TEMs in the absence of impurity pressure gradients, do not match the sign. In ref.~\cite{McKinney_jpp_2019} the ITG-driven normalized heat transport for HSX is significantly lower than for NCSX, despite the larger linear growth rate exhibited by HSX}. Finally, regarding the impurity transport under different types of background turbulence, ITG turbulence is known to drive impurity radial flux more efficiently than TEM turbulence \cite{Regana_JPP_2021} in W7-X (when driven solely by the ion temperature and density gradient, respectively). The calculations that follow now assess the transport coefficients in the presence of a mixed type of turbulence in W7-X only, when the density gradient of the main species is gradually increased and the ion temperature gradient is kept constant.

\begin{figure}
	\includegraphics[width=0.5\textwidth]{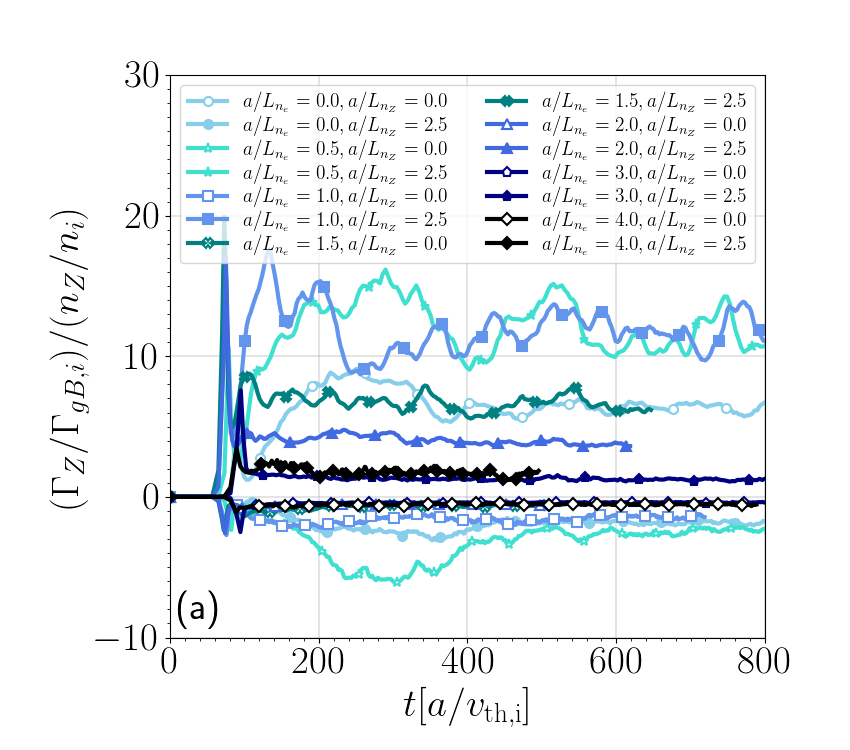}
	\includegraphics[width=0.5\textwidth]{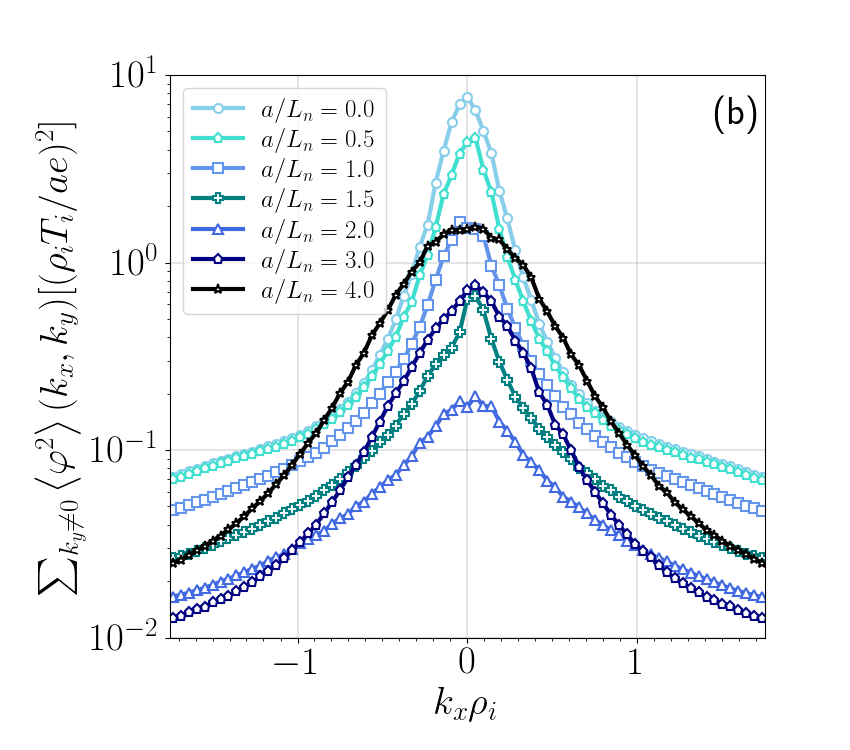}
	\caption{\blue{For Fe$^{16+}$ impurities at trace concentration with $a/L_{T_Z}=a/L_{T_i}=4.0$ at $r/a=0.75$: (a) time trace of the normalised radial particle flux for the set of simulations with vanishing impurity density gradient (open symbols) and finite impurity density gradient (filled symbols); (b) for the different normalised density gradients, $k_x$-spectra of the non-zonal components of the line-averaged turbulent electrostatic potential squared.}}
	\label{fig:w7x_nprim_scan_sims}	
\end{figure}

\begin{figure}
	\begin{center}
		\includegraphics[width=0.7\textwidth]{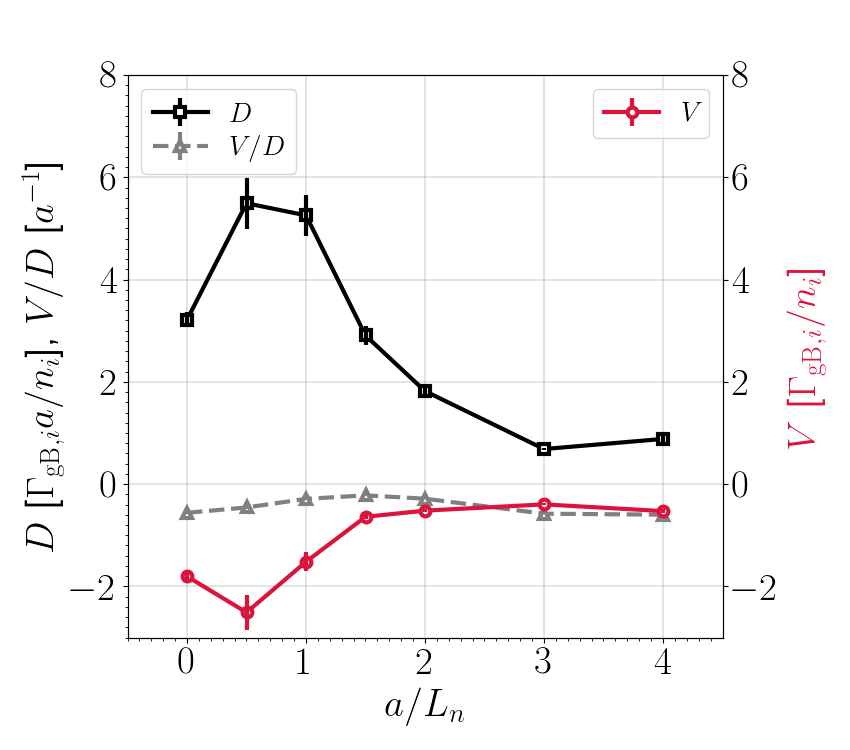}
		\caption{\blue{For Fe$^{16+}$ impurities at trace concentration with $a/L_{T_Z}=a/L_{T_i}=4.0$ at $r/a=0.75$, normalised diffusion coefficient (squares), convection coefficient (circles) and turbulent peaking factor (triangles) as a function of the normalised density gradient of the main ions and electrons, $a/L_n$.}}
		\label{fig:w7x_nprim_scan_D_and_V}
	\end{center}
\end{figure}

The nonlinear simulations performed for this purpose have considered, apart from kinetic ions and electrons, Fe$^{16+}$ at \refone{a trace concentration of $n_Z/n_i=10^{-12}$}. The temperature gradient of the impurity has been set equal to that for the ions, $a/L_{T_Z}=a/L_{T_i}=4.0$, and, as in the previous sections, for each value of the density gradient for the main species, two different values of the impurity density gradient have been considered, in order to calculate the coefficients $V$ and $D$. The electron temperature gradient has been set to zero in order to preclude the contribution from $T'_e$-driven turbulence. The only difference with respect to the settings of the nonlinear simulations presented in section \ref{sec:transport} has been the resolution chosen for the simulations with finite density gradient\refone{, which has also covered a slightly different set of values: $a/L_n=\left\{0.0, 0.5, 1.0, 1.5, 2.0, 3.0, 4.0\right\}$}. In order to guarantee that the spectrum of the turbulent electrostatic potential, which tends to spread along $k_x$ as the density gradient increases, is well captured within the Fourier box, and that the particle trapping is well resolved when the transition from pure ITG to a mixed turbulence with contribution from TEMs occurs, the resolution chosen has been set to $\{N_x, N_y, N_z, N_{v_{\|}}, N_{\mu}\}=\{114,164,96,36,12\}$. The time traces of the normalised impurity particle flux from the resulting \refone{fourteen} simulations are shown in figure \ref{fig:w7x_nprim_scan_sims} (a). From that figure, one can immediately infer the negative sign of the convection and the comparatively larger diffusive contribution to the total transport. In figure \ref{fig:w7x_nprim_scan_sims} (b) the $k_x$-spectrum of the line averaged turbulent electrostatic potential squared is represented and, as just mentioned, it can be appreciated how, as the density gradient of the main species increases, the maximum amplitude of the turbulent potential decreases and becomes more extended along $k_x$. That trend reverses as the density gradient varies from $a/L_n=2.0$ to $a/L_n=4.0$ and the maximum amplitude of the potential increases again.
From the time averaged value of the impurity fluxes obtained during the saturated phase, the diffusion and convection coefficients as well as the peaking factor, shown in \refone{figure ~\ref{fig:w7x_nprim_scan_D_and_V}}, are finally calculated. The convection velocity remains inwards throughout the scan. \refone{It first increases with the density gradient, from $a/L_n=0$ to $a/L_n=0.5$, and then decreases uniformly up to a factor of approximately $6$ in the range between $a/L_n=0.5$ and $a/L_n=3.0$, from $V\sim 2.5$} to $V\sim 0.4$ (in normalised units). A slight increase follows at $a/L_n=4.0$. With regard to the diffusion coefficient, its dependence with $a/L_n$ \refone{shows a significant increase from $a/L_n=0$ to $a/L_n=0.5$} followed by an abrupt fall of nearly a factor of $8$ reached at $a/L_n=3.0$. The resulting turbulent peaking factor \revone{takes} rather low values in the range from $V/D\sim -0.6$ to $V/D\sim -0.3$.\\
Considering the plasma parameters in the different W7-X scenarios \cite{Carralero_nf_submitted_2021}, these results \reftwo{are consistent with the following picture}. In the frequently performed ECRH W7-X discharges, weak density gradients of the main species (typically $a/L_n\lesssim 1$) develop at the edge and, in that case, the transport of impurities will be dominated by a strong turbulent diffusion, which will efficiently counteract any source of inward convective transport, both neoclassical (under ion-root conditions) and turbulent. Such balance of diffusion and convection will lead to a weak impurity density peaking. On the other hand, the enhanced pellet fueled confinement scenarios, which increase the density peaking of the main species to typically $2\lesssim a/L_n\lesssim 3$, will reduce the diffusion coefficient drastically as well as the convective turbulent source. It is important to note that, even under these conditions, the turbulent transport of impurities is not suppressed and, indeed, the turbulent diffusion will still be larger than the neoclassical diffusion, as the former decreases a large factor but not orders of magnitude \cite{Geiger_NF_59_046009_2019}. However, the decrease of both turbulent transport coefficients would obviously make the \refone{collisional channels of transport, both classical, which can be eventually large in W7-X \cite{Buller2019}, and} neoclassical gain greater relative weight on the transport of impurities. In particular, the more prominent role of the neoclassical convection, together with the enhancement of the ambipolar ion-root at the edge of these plasma \cite{Estrada_nf_2021}, could qualitatively explain the observed trends of the longer impurity confinement time and the observation of radial localisation of impurities in high performance scenarios. 



\section{Discussion}
\label{sec:discussion}
Turbulence has been pointed out as a prominent player in the transport of impurities during the first campaigns of W7-X and, specifically, in preventing their accumulation. Although experiments have collected \revone{evidence that supports} the role of turbulence on the transport of impurities, it remains unclear whether the underlying properties of the turbulent impurity transport in W7-X, to a large extent validated against numerical simulations \cite{Regana_JPP_2021}, are specific of that device or common to other stellarators. This has been \blue{one of the two} questions we have addressed in this contribution. Based on nonlinear gyrokinetic simulations performed \blue{with the code \stella} for W7-X, LHD, TJ-II and NCSX, we have concluded that there is nothing essentially different among W7-X, TJ-II and NCSX with regard to impurity transport \blue{driven by pure ITG turbulence}. Only LHD deviates appreciably for most of the results shown here from the behavior exhibited by the other three stellarators.
In particular, the diffusive transport contribution in W7-X is not actually stronger than in TJ-II or NCSX, and although the size of $D$ differs between them, so does the convection velocity such that the turbulent peaking factor is roughly the same for the three devices. It is important to remark that the value of the peaking factor, although negative, is fairly low for these three stellarators and, at worst, rather moderate for LHD. Thus, it does not seem that ITG \revone{turbulence} can produce strong \blue{impurity density peaking} in stellarators. \blue{Nonetheless}, \blue{moderate to large} impurity density gradients and longer confinement times have been measured in W7-X scenarios of reduced turbulence \cite{Langenberg_IAEA_2021, Stechow_submitted_2021}. Thus, the second question addressed in the present contribution has focused on W7-X and, in particular, on how the increase of the density peaking of the main species, specific of pellet fueled high performance scenarios, affects the turbulent transport of impurities. It has been numerically shown that under such conditions of mixed ITG/TEM turbulence, both the convection coefficient and, even in more degree,  the diffusion coefficient can be significantly reduced. This behavior of the transport coefficients, which would lead to a larger specific weight of the neoclassical convection, aligns qualitatively well with the few experimental evidences that show an increase of the impurity density peaking and confinement time when turbulence is reduced.

\section*{Acknowledgements}
This work has been carried out within the framework of the EUROfusion Consortium and has received funding from the Euratom research and training programme 2014-2018 and 2019-2020 under grant agreement No. 633053. The views and opinions expressed herein do not necessarily reflect those of the European Commission. This research was supported in part by grant PGC2018-095307-B-I00, Ministerio de Ciencia, Innovaci\'on y Universidades, Spain. The simulations were carried out in the clusters Marconi (Cineca, Italy) and Xula (Ciemat, Spain). 

\section*{Bibliography}


\end{document}